\documentstyle[epsf,12pt]{article}
\newcommand{\yourabstract}[1]{
            \mbox{}\\
            \mbox{}\\
            {\bf\noindent Abstract}\\
            \begin{center}
            \mbox{}\parbox[t]{5.in}{#1}
            \end{center} }

\begin{document}
\begin{titlepage}
\begin{flushright}
CERN-TH.7297/94 \\
hep-th/9406085
\end{flushright}
\vskip 1.5cm
\begin{center}
{\bf \large Entropy, Quantum Decoherence and Pointer States}
\vskip 0.25cm
{\bf \large in Scalar ``Parton'' Fields}
           \footnote{
           Work supported by the Heisenberg
           Programme (Deutsche Forschungsgemeinschaft).
           }
\vskip 0.8cm
Hans-Thomas Elze
\vskip 0.25cm
CERN-Theory,
CH-1211 Geneva 23,
Switzerland \footnote{ELZE@CERNVM.CERN.CH}
\end{center}
\yourabstract{Entropy arises in strong interactions
by a dynamical separation of
``partons'' from unobservable
``environment'' modes due to
confinement. For interacting scalar fields we
calculate the statistical entropy
of the observable subsystem. Diagonalizing
its density matrix
yields field pointer states and
their probabilities in terms of Wightman
functions. It also indicates how to calculate a finite
geometric entropy proportional to a
surface area.
}
\vskip 0.15cm
\begin{center}
\vskip 0.6cm
{\it Submitted to Physical Review Letters}
\end{center}
\vskip 0.4cm
\flushleft{CERN-TH.7297/94

June 1994}
\end{titlepage}

The long-standing ``entropy puzzle'' of
high-multiplicity events in strong interactions at high energy has
been analysed from a new point of view \cite{I}. The problem dates
back to Fermi and Landau
and is related to understanding the
rapid thermalization of high energy density ($\gg 1\;\mbox{GeV/fm}^3$)
matter \cite{Fermi}. Why do thermal models work so well? Why do they
work at all?

Or, why does high-energy
scattering of pure initial states lend itself to a
statistical description characterized by large apparent entropy
from a mixed-state density matrix describing
intermediate stages in a space-time picture of parton evolution?
Effectively, {\it unitary time evolution} of the observable part
of the system breaks down in the
transition from a quantum mechanically pure initial state to a
highly impure (more or less thermal) high-multiplicity final state.
In Ref. \cite{I} this was discussed in detail. Based on
analogies with studies of the quantum measurement process
(``collapse of the wave function'') \cite{Zu0} and
motivated by related problems in quantum cosmology and by
non-unitary non-equilibrium evolution
resulting in string theory \cite{GH}, we argued that
{\it environment-induced quantum decoherence solves the
entropy puzzle} of strong interactions.

A complex pure-state quantum system can show
quasi-classical behaviour, i.e. an impure density (sub)matrix
together with decoherence of associated {\it pointer states}
in the observable subsystem \cite{I}. In particular, there is
a {\it Momentum Space Mode Separation}
due to confinement, which is defined in
the frame of initial conditions for the time evolution and for
the physical (gauge) field degrees of freedom. Thus, almost constant
QCD field configurations form an {\it unobservable environment},
since they neither hadronize nor initiate hard scattering among
themselves. It interacts with the {\it observable subsystem} composed
of partons \cite{P}.

The induced quantum decoherence and entropy production were
studied in a non-relativistic single-particle model resembling an
electron coupled to the quantized electromagnetic field,
however, with an enhanced oscillator spectral density
in the infrared. The Feynman-Vernon influence
functional technique for quantum Brownian motion
\cite{PHZ} provided the remarkable result that
in the {\it short-time strong-coupling limit} the model parton
behaves like a {\it classical particle} \cite{I}:
Gaussian parton
wave packets experience {\it friction} and {\it localization}, i.e.
no quantum mechanical spreading, and their coherent {\it superpositions
decohere}.

Summarizing,
partons feel an unobservable
(gluonic) environment, which manifests its strong non-perturbative
interactions
on a short time scale ($\ll 1\;$fm/{\it c}) through
{\it decoherence} of partonic pointer states, their
{\it quasi-classical behaviour}, and {\it entropy production}.
If confirmed in QCD, this will have important consequences
for parton-model applications to complex hadronic
or nuclear reactions \cite{GM}. The emergence of
{\it structure functions} from initial-state wave functions
will be further studied in our approach.

We defined a model of two coupled scalar fields
representing partons and their non-perturbative environment.
In the functional
Schr\"odinger picture employing Dirac's time-dependent variational
principle
we derived its Cornwall-Jackiw-Tomboulis (CJT) effective action and the
equations of motion for renormalizable interactions \cite{I,Corn}. Thus,
analysis of the entropy puzzle in strong interactions
leads to study an observable field
(open subsystem) interacting with a dynamically hidden one
(unobservable environment), i.e. {\it quantum field Brownian motion}.

In the following we
derive the {\it entropy in any system of two
interacting real scalar fields}.
Their most general normalized {\it Gaussian
wave functional} in the Schr\"odinger picture
can be written as
\begin{eqnarray}
\Psi _{12}[\phi _1,\phi _2;t]&\equiv&N_{12}(t)\;\Psi _{G_1}[\phi _1;t]\;
\Psi _{G_2}[\phi _2;t]
\nonumber \\ [2ex]
&\;&\cdot\;
\exp\left\{-\textstyle{\frac{1}{2}}
[\phi _1-\bar{\phi} _1(t)]\left [G_{12}(t)-i
\Sigma _{12}(t)\right ]
[\phi _2-\bar{\phi} _2(t)]\right\}
\;, \label{1} \end{eqnarray}
with ($j=1,2$)
\begin{eqnarray}
&\;&\Psi _{G_j}[\phi _j ;t]\;\equiv \label{2} \\ [2ex]
&\;&N_j(t)\;\exp \left\{
-[\phi _j-\bar{\phi}_j(t)]
\left [\textstyle{\frac{1}{4}} G^{-1}_j(t)
-i\Sigma _j(t)\right ]
[\phi _j-\bar{\phi}_j(t)]\; +\; i
\bar{\pi}_j(t)
[\phi _j-\bar{\phi}_j(t)] \right\}
\;. \nonumber \label{200} \end{eqnarray}
We suppress all spatial integrations.
The normalization factors are
\begin{equation}
N_j(t)\; =\;\det \{ 2G_j(t)\}^{-1/4}
\;\; ,\;\;\;
N_{12}(t)\; =\;
\det \{ 1-G_1(t)G_{12}(t)G_2(t)G_{12}(t)\}^{1/4}
\;\;, \label{3} \end{equation}
discarding an irrelevant constant factor in $N_j$.
Thus, the time-dependent Hartree-Fock approximation
(TDHF) for the quantum field
Schr\"odinger equation \cite{I,Corn} is embodied in the
variational parameter
one-point functions $\bar{\phi}_j(x,t)$, $\bar{\pi}_j
(x,t)$ (mean fields) and symmetric two-point functions
$G_j(x,y,t)$, $\Sigma _j(x,y,t)$, $G_{12}(x,y,t)$, $\Sigma _{12}(x,y,t)$
 (related to Wightman functions).
Their meaning was discussed in
\cite{I}.

All physical quantities of the complex system can be calculated with
$\Psi _{12}$, expressing inner products by functional
integrals. The functional density submatrix
$\hat{\rho}_{\cal P}$ for the observable ``parton'' subsystem
($\phi _1$) is obtained by tracing over the unobservable degrees of
freedom ($\phi _2$),
\begin{equation}
\hat{\rho} _{\cal P}(t)\;\equiv\;
\mbox{Tr} _2\; |\Psi _{12}(t)\rangle\langle\Psi _{12}(t)|
\;\;, \label{4} \end{equation}
as calculated explicitly in \cite{I} (we henceforth omit ${\cal P}$).
The matrix elements of $\hat{\rho}$
contain {\it all} the information about the subsystem.
Our aim is to obtain the {\it von Neumann} or {\it statistical
entropy}, $S\equiv -\mbox{Tr} _1\;
\hat{\rho}\ln\hat{\rho}$.
Before, we calculated the simpler {\it linear entropy}
directly, which provides a lower bound for the statistical
entropy \cite{I},
\begin{equation}
S(t)\;\geq\;
-\frac{1}{2}\;\mbox{Tr}
\;\ln\left (\frac{\textstyle{1-G_1(t)G_{12}(t)G_2(t)G_{12}(t)}}
{\textstyle{1+G_1(t)\Sigma _{12}(t)G_2(t)\Sigma _{12}(t)}}\right )
\;\;, \label{5} \end{equation}
tracing over coordinates.
Equation (\ref{5}) is also valid for {\it non-translation
invariant} systems, which is relevant for calculating the
geometric entropy related to spatial
boundaries separating observable and unobservable subsystems.

{\it Geometric entropy} is intimately connected to black-hole
entropy \cite{Sred}.
Here, one identifies $\phi _1$ as the part of a
scalar field $\phi$ with support {\it outside} a given spatial region
and $\phi _2\equiv\phi -\phi _1$, which has its support {\it inside}
the complement.
Our results (\ref{5})
and (\ref{17}) below indicate that geometric entropy
comes out {\it finite}, once a renormalization of the equations
for the two-point functions, $G$'s and $\Sigma$'s in
(\ref{5}), is performed or a UV regularization introduced to provide
sufficient integrability constraints.

We proceed by {\it diagonalizing} $\hat{\rho}$.
Determining its eigenstates and eigenvalues is equivalent
to constructing {\it field pointer states}
\cite{I,Zu0} within TDHF and their probabilities.
The eigenvalue problem $\hat{\rho}|\rho\rangle =
\rho |\rho\rangle$ to be solved is of the form
\begin{eqnarray}
&\;&\rho\;
F[\phi ]\;\exp\{-
\phi\alpha\phi +\beta\phi\}\; =\; \label{6} \\ [2ex]
&\;&N^2\;\exp\{ -\phi a\phi +b\phi\}\int {\cal D}\phi '\; F[\phi ']\;
\exp\{ -\phi '[a^{\ast}+\alpha ]\phi '+[b^{\ast}+\beta +\phi c^{\ast}]
\phi '\}
\;\;, \nonumber \label{600} \end{eqnarray}
using the ansatz $\langle (\phi +\bar{\phi} _1)|\rho\rangle\equiv
F[\phi ]\exp\{ -\phi\alpha\phi +\beta\phi\}$ with unknown one-
and two-point functions $\beta$ and $\alpha$ and a non-exponential
functional $F$.
According to results for
$\hat{\rho}_{\cal P}$ from \cite{I}, we define
$N\equiv N_1N_{12}$, $b\equiv i\bar{\pi} _1$ ($\bar{\phi} _1$
does not appear in (\ref{6})),
\begin{eqnarray}
a&\equiv&\textstyle{\frac{1}{4}} G_1^{-1}A
-i[\Sigma _1-\textstyle{\frac{1}{8}}(G_{12}G_2\Sigma _{12}
+\Sigma _{12}G_2G_{12})]\; =\; a^t\;\;, \nonumber \\
c&\equiv&\textstyle{\frac{1}{2}} G_1^{-1}B
-\textstyle{\frac{i}{4}}[G_{12}G_2\Sigma _{12}-\Sigma _{12}G_2G_{12}]
\; =\; c^{\dagger}
\;\;, \nonumber \label{601} \end{eqnarray}
and combinations of two-point functions
\begin{eqnarray}
A&\equiv &1-\textstyle{\frac{1}{2}} G_1G_{12}G_2G_{12}
+\textstyle{\frac{1}{2}} G_1\Sigma _{12}G_2\Sigma _{12}
\;\;, \nonumber \\
B&\equiv&\textstyle{\frac{1}{2}} G_1G_{12}G_2G_{12}
+\textstyle{\frac{1}{2}}
G_1\Sigma _{12}G_2\Sigma _{12}
\;\;. \nonumber \label{602} \end{eqnarray}
Choosing $\beta\equiv b$ in (\ref{6}), completing the square,
shifting $\phi '$, and requiring resulting Gaussians in $\phi$
to cancel yields the eigenvalue
problem:
\begin{equation}
\rho\; F[\phi ]\; =\; N^2\int {\cal D}\phi '\; F[\phi '+Y\phi ]\;
\exp\{ -\phi 'X\phi '\}
\;\;, \label{7} \end{equation}
with $X\equiv a^{\ast}+\alpha =X^t$, $Y\equiv\textstyle{\frac{1}{2}}
X^{-1}c$, and where
$\alpha =\alpha ^t$, by (\ref{6}), is determined to solve the equation
$a-\alpha =\textstyle{\frac{1}{4}} c^{\ast}[a^{\ast}+\alpha ]^{-1}c$.
Note the similarity to the
finite-dimensional oscillator problem of Srednicki
\cite{Sred}.

Equivalently, replacing
$F[\phi ']\rightarrow
F[\delta /\delta (\phi c^{\ast})+\delta /\delta (c\phi )]$ and
$\phi c^{\ast}\phi '\rightarrow\textstyle{\frac{1}{2}}
[\phi c^{\ast}\phi '+\phi 'c\phi ]$
in (\ref{6}), we obtain by integration
\begin{equation}
\rho\; F[\phi ]\; =\; N^2\;\det \{ X\}^{-1/2}\;\exp\{
-\textstyle{\frac{1}{4}}\phi c^{\ast}X^{-1}c\phi\}\;
F[\frac{\delta}{\delta (\phi c^{\ast})}+\frac{\delta}{\delta (c\phi )}]
\;\exp\{
+\textstyle{\frac{1}{4}}\phi c^{\ast}X^{-1}c\phi\}
\;\;, \label{8} \end{equation}
which is more convenient than (\ref{7}).
Looking for polynomial functional solutions of (\ref{8}), we find
first of all a {\it constant},
\begin{equation}
F_0[\phi ]\;\equiv\; 1\;\;\;\Rightarrow\;\;\rho _0\; =\; N^2\;\det \{
X\} ^{-1/2}
\;\;. \label{9} \end{equation}
Secondly, instead of a general {\it linear} functional,
the Fourier transform is sufficient,
\begin{equation}
F_k[\phi ]\;\equiv\;\int d^dx\;\mbox{e} ^{-ikx}\phi (x)\;\equiv\;
\phi _k
\;\;, \label{10} \end{equation}
since the problem is linear in $F$. Then, from (\ref{8}) $-$ (\ref{10}),
\begin{equation}
\rho _1\phi _k\; =\;\textstyle{\frac{1}{4}}\rho _0
[X^{-1}c\phi +\phi c^{\ast}X^{-1}]_k\;
=\;\rho _0Y_{kk'}\phi _{-k'}
\;\;, \label{11} \end{equation}
summing over indices occurring twice. For a translation-invariant
system, (\ref{11}) could immediately be solved. Generally, however,
denoting
eigenvalues and eigenvectors of $(Y_{kk'})$ by $\xi _k$ and
$\tilde{\phi} _k$, one obtains a set of {\it linear} eigenvalues
$\rho _k=\rho _0\xi _k$. Due to the Gaussian structure in (\ref{8}),
the {\it higher-order eigenfunctionals} can be built up as linear
combinations of products of $\tilde{\phi} _k$'s and lower-order ones.
For example,
$F_{kk'}[\phi ]\equiv \tilde{\phi} _k\tilde{\phi} _{k'}
+C_{kk'}$, which yields a set of {\it quadratic} eigenvalues
$\rho _{kk'}=\rho _0\xi _k\xi _{k'}\Theta (k'-k)$. Note the constraint
$k'\geq k$; interchange of $k$ and $k'$ does not lead to a new
eigenfunctional due to the scalar (bosonic) character of the fields.
The {\it constant} $C_{kk'}$ follows with the help of the matrix
diagonalizing $(Y_{kk'})$. We do not construct explicitly the
higher-order eigenfunctionals. However, the {\it n-th order set
of eigenvalues},
\begin{equation}
\rho _{k_1\ldots k_n}\; =\;\rho _0\xi _{k_1}\prod _{i=2}^{n}
\;\xi _{k_i}\Theta (k_i-k_{i-1})
\;\;, \label{12} \end{equation}
is easily found, similarly to $\rho _{kk'}$ above.
To check the result (\ref{12}), we calculate
\begin{eqnarray}
\mbox{Tr}\;\hat{\rho} (t)&=&\sum \;\mbox{eigenvalues}\;
=\;\rho _0+\sum _{n=1}^\infty\sum _{k_1\ldots k_n}\;\rho _{k_1\ldots
k_n} \nonumber \\
&=&\rho _0[1+\sum _{k_1}\;\xi _{k_1}+\sum _{k_1\leq k_2}\;
\xi _{k_1}\xi _{k_2}+\sum _{k_1\leq k_2\leq k_3}\;\xi _{k_1}\xi _{k_2}
\xi _{k_3}+\;\ldots\; ] \nonumber \\
&=&\rho _0\prod _k\sum _{n_k=0}^\infty\;\xi _k^{\; n_k}\; =\;
\rho _0\prod _k\; [1-\xi _k]^{-1}\; =\;\rho _0\det \{1-Y\} ^{-1}\;
=\; 1
\;\;, \label{13} \end{eqnarray}
which resembles the evaluation of a bosonic partition function. In
the last step we used $\rho _0=\det\{ X^{-1}\mbox{Re}[2a-c]\} ^{1/2}=
\det\{ [1-\textstyle{\frac{1}{2}} X^{-1}c^{\ast}]
[1-\textstyle{\frac{1}{2}} X^{-1}c]\} ^{1/2}$, which
follows from the equation determining $\alpha$ or $X$.

Similarly, we obtain the {\it linear entropy},
\begin{equation}
S ^{lin}\;\equiv\;\mbox{Tr} _1\left\{
\hat{\rho}-\hat{\rho} ^{\; 2}\right\} \; =\;
1-\;\mbox{Tr} _1\;\hat{\rho} ^{\; 2}\; =\;
1-\;\det\left\{\frac{1-Y}{1+Y}\right\}
\;\;. \label{14} \end{equation}
In order to express $Y$ in terms of $A$ and $B$, we observe that
in a {\it direct} calculation \cite{I} of $\mbox{Tr}_1\hat{\rho}^2$
(and in the {\it n}-fold functional integral for
$\mbox{Tr}_1\hat{\rho}^n$)
{\it imaginary parts} of $a$ and
$c$ cancel. Therefore, we replace $a$ and $c$ by their {\it real parts},
$a=\textstyle{\frac{1}{4}} G_1^{-1}A$ and
$c=\textstyle{\frac{1}{2}} G_1^{-1}B$, simplifying the
equation for $\alpha$, $X$, or $Y$, $a-\alpha =\textstyle{\frac{1}{4}} c
[a+\alpha ]^{-1}c$. The solution (for integrable
eigenfunctionals) is
\begin{equation}
Y\; =\; (c/2)^{-1/2}\left [\tilde{a}+(\tilde{a}^2-1)^{1/2}\right ]
^{-1}(c/2)^{1/2}
\;\;, \label{15} \end{equation}
with $\tilde{a}\equiv (c/2)^{-1/2}a(c/2)^{-1/2}$.
Finally, inserting (\ref{15}) into (\ref{14}),
\begin{equation}
S^{lin}(t)\; =\; 1-\;\det\left\{\frac{A(t)-B(t)}{A(t)+B(t)}\right\}
^{1/2}
\;\;, \label{16} \end{equation}
which confirms our earlier result, employed in
(\ref{5}) \cite{I}.
Next, we calculate the {\it statistical entropy}
using the ``replica trick'':
\begin{eqnarray}
S(t)&\equiv&-\;\mbox{Tr} _1\;\hat{\rho}(t)\ln\hat{\rho}(t)\;
=\; -\frac{d}{dn}\mbox{Tr} _1\;\hat{\rho}^{\; n}\; |_{n=1}\;
=\; -\frac{d}{dn}\det\left\{\frac{(1-Y)^n}{1-Y^n}\right\}|_{n=1}
\nonumber \\
&=&-\;\mbox{Tr}\left\{\ln (1-Y)+\frac{Y}{1-Y}\ln Y\right\}
\;\;. \label{17} \end{eqnarray}
Together with (\ref{15}), eq. (\ref{17}) presents our main result.
It generalizes eq. (\ref{6})
of Srednicki \cite{Sred}. Basically,
the TDHF approximation for {\it interacting} quantum fields
preserves a Gaussian structure of the wave functionals,
see (1) $-$ (3), which is exact in the non-interacting case and
can be reduced to a coupled harmonic oscillator problem.

To evaluate the entropy (\ref{17}) is still a
formidable task for any realistic situation. Before trying,
it seems worth while to draw some general conclusions:
\vskip .15cm
\noindent
{\bf I.} Neither mean fields $\bar{\phi} _{1,2}$, nor their
conjugate momenta $\bar{\pi} _{1,2}$, nor imaginary parts
$\Sigma _{1,2}$ of the ``parton'' and environment two-point
functions
contribute to $S$.
\vskip .15cm
\noindent
{\bf II.} {\it Vanishing correlations
between ``partons'' and environment}, i.e. $G_{12}=\Sigma _
{12}=0$ (independent subsystems), {\it imply} $A=1,B=0$, i.e. $Y=0$,
{\it and} $S=0$.
\vskip .15cm
\noindent
{\bf III.} {\it Vanishing widths of ``parton'' or environment
wave functionals}, i.e.
$G_{1,2}\rightarrow 0$
(one or the other subsystem
classical/reversible \cite{I}),
{\it imply} $Y=0$ {\it and} $S=0$.
\vskip .15cm
\noindent
This presumably holds for {\it any} field theory of ``partons''
coupled to environment modes independently of the
interactions in TDHF approximation.
The time-evolution, however, follows specific
equations of motion for the one- and two-point functions \cite{I}.

Our considerations confirm
that {\it quantum decoherence}
and {\it entropy production} in a subsystem is
induced by an active
environment \cite{I,Zu0,GH,PHZ}. The above diagonalization
of the ``parton'' density functional
yields time-dependent {\it field pointer states},
the simplest one of largest probability $\rho _0$ being
\begin{equation}
\Psi _0[\phi ;t]\; =\;\exp\{ -[\phi -\bar{\phi}_1(t)]\alpha (t)
[\phi -\bar{\phi}_1(t)]+i\bar{\pi}_1(t)[\phi -\bar{\phi}_1(t)]\}
\;\;, \label{18} \end{equation}
cf. (\ref{6}) $-$ (\ref{9}), with $\alpha = [(c/2)(\tilde{a}^2-1)(c/2)]
^{1/2}$.
Higher-order eigenfunctionals are less
probable, see ({\ref{12}), {\it and} have higher kinetic energy, since
their wave functionals have additional nodes, e.g. (\ref{10}),
analogous to excited oscillator states.

As a first application of (\ref{17})
we consider the {\it large-entropy limit}, i.e.
$Y\approx 1$ or $A\approx B$. Then, using (\ref{14}) and (\ref{16}),
we find:
\begin{eqnarray}
S(t)&\approx&
-\;\mbox{Tr}\;\ln (1-Y)/(1+Y)\; =\;
-\textstyle{\frac{1}{2}}\;\mbox{Tr}\;\ln (A-B)/(A+B) \nonumber \\
&=&\textstyle{\frac{1}{2}}
\sum _{n=1}^\infty \frac{1}{n}\;\mbox{Tr}\;\{ [G_1G_{12}G_2G_{12}]^n
-[-G_1\Sigma _{12}G_2\Sigma _{12}]^n\}
\;\;, \label{19} \end{eqnarray}
i.e. (\ref{5}).
If we assume a {\it spatial surface} of area ${\cal A}$
dividing the system into two,
which is flat on the scale of the {\it short-ranged}
correlations in (\ref{19}),
then $\mbox{Tr}[\ldots ]^n$ can be interpreted as a sum of closed loops
of strings of $G$'s or $\Sigma$'s
intersecting the surface $2n$ times: once for each factor $G_{12}$ or
$\Sigma _{12}$ correlating in- and outside fields. The
dominant contribution to the trace comes from small loops (let
$\Sigma _{12}^{\;2}\ll G_{12}^{\;2}$).
Regularizing their contribution by a short-distance cut-off
${\cal L}$,
their size transverse to the surface will be
O(${\cal L}^2$) for $d=3$. Transverse to the surface the system
is locally translation-invariant. Therefore, the {\it geometric
entropy} is approximately
\begin{equation}
S(t)\;\propto\;-\frac{{\cal A}}{{\cal L}^2}\;\mbox{Tr}_{\cal L}
\;\ln\left (\frac{1-G_1G_{12}G_2G_{12}}{1+G_1\Sigma _{12}G_2\Sigma _{12}}
\right )
\;\;, \label{20} \end{equation}
where $\mbox{Tr}_{\cal L}$ is evaluated locally on the scale of
${\cal L}$. A dimensional analysis led Srednicki to
propose $S\propto {\cal A}$
before \cite{Sred}. Equation (\ref{20}) can be applied to the
moving mirror model; following Kabat et al. \cite{Sred}, it
approximates the {\it thermal entropy} outside a black hole of
radius much larger than ${\cal L}$.

Secondly, coming back to partons interacting with their (gluonic)
environment, the {\it rate of entropy production}, which follows
from (\ref{17}), is most interesting. We define a dynamical {\it
decoherence time} $\tau$,
\begin{equation}
\tau ^{-1}(t)\;\equiv\;\frac{d}{dt}\;\ln\; S(t)\;\approx\;
\frac{\mbox{Tr}\;\dot{Y}\ln Y}{\mbox{Tr}\; Y\ln Y}\; =
\;\frac{\int\tilde{d}k\;\dot{Y} _k\ln Y_k}{\int\tilde{d}k\; Y_k\ln Y_k}
\;\;, \label{21} \end{equation}
with $\tilde{d}k\equiv d^dk/(2\pi )^d$.
For simplicity we assumed small $Y$ or $S$ and a translation- invariant
system; the Fourier transform is
$Y_k=B_k[A_k+(A_k^{\; 2}-B_k^{\; 2})^{1/2}]^{-1}$, since $A,B$ are
convolutions of two-point functions now. Generally, two limits are
particularly important: $\tau (t\rightarrow 0)$ gives the time scale
for the decay of a Gaussian partonic field state, cf. (\ref{1}) $-$
(\ref{3}), into an {\it incoherent superposition} of pointer states,
e.g. (\ref{18}), with impure density matrix and {\it non-zero entropy};
$\tau (t\gg 0)$ reflects the approach to a stationary state (\it
thermalization}), if it exists. Using the equations of motion from
\cite{I}, the decoherence time will be calculated for phenomenologically
 interesting situations elsewhere.
\vskip .15cm
\noindent
I thank N. E. Mavromatos, L. Pesce, and J. Rafelski for
stimulating discussions.


\begin{thebibliography}{15}
\bibitem{P} They are observable in the sense of parton-hadron
            duality or deep-inelastic scattering.
\end{thebibliography}
\end{document}